\documentclass[structabstract]{aa}  
\usepackage{graphicx}
\usepackage{txfonts}
\usepackage{natbib}
\usepackage{color}
\usepackage{lscape}
\usepackage{latexsym}
\bibpunct{(}{)}{;}{a}{}{,} 
\begin{document}
\definecolor{Red}{rgb}{1,0,0}

   \title{A Census of Stellar Mass in 10 Massive Haloes at $z\sim 1$ from the GCLASS Survey}
   \author{Remco F.J. van der Burg\inst{1}
          \and Adam Muzzin\inst{1}
          \and Henk Hoekstra\inst{1} 
          \and Gillian Wilson\inst{2}
          \and Chris Lidman\inst{3}
          \and H.K.C. Yee\inst{4}}
	
   \institute{Leiden Observatory, Leiden University, P.O. Box 9513, 2300 RA Leiden, The Netherlands\\
                 \email{vdburg@strw.leidenuniv.nl}        
             \and Department of Physics and Astronomy, University of California-Riverside, 900 University Avenue, Riverside, CA 92521, USA
             \and Australian Astronomical Observatory, PO Box 915, North Ryde, NSW 1670, Australia
             \and Department of Astronomy \& Astrophysics, University of Toronto, Toronto, Ontario M5S 3H4, Canada
             }             
             
   \date{Received ...; accepted ...}
  \abstract {}
    {We study the stellar mass content of massive haloes in the redshift range $0.86 < z < 1.34$, by measuring: (1) The stellar mass in the central galaxy versus total dynamical halo mass. (2) The total stellar mass (including satellites) versus total halo mass. (3) The radial stellar mass and number density profiles for the ensemble halo.}
    {We use a K$_{\mathrm{s}}$-band selected catalogue for the 10 clusters in the Gemini Cluster Astrophysics Spectroscopic Survey (GCLASS), with photometric redshifts and stellar masses measured from 11-band SED fitting. Combining the photometric catalogues with the deep spectroscopic component of GCLASS, we correct the cluster galaxy sample for interlopers. We also perform a dynamical analysis of the cluster galaxies to estimate the halo mass $M_{200}$ for each cluster based on a measurement of its velocity dispersion.}
    {(1) We find that the central galaxy stellar mass fraction decreases with total halo mass, and that this is in reasonable quantitative agreement with measurements from abundance matching studies at $z\sim 1$. 
     (2) The total stellar mass fractions of these systems decrease with halo mass, indicating that lower mass systems are more efficient at transforming baryons into stars. We find the total stellar mass to be a good proxy for total halo mass, with a small intrinsic scatter. When we compare these results from GCLASS with literature measurements, we find that the stellar mass fraction at fixed halo mass shows no significant evolution in the range $0 < z < 1$.
     (3) We measure a relatively high NFW concentration parameter $\rm{c_{g}} \sim 7$ for the stellar mass distribution in these clusters, and debate a possible scenario to explain the evolution of the stellar mass distribution from the GCLASS sample to their likely descendants at lower redshift.}
    {The stellar mass measurements in the $z\sim 1$ haloes provided by GCLASS puts constraints on the stellar mass assembly history of clusters observed in the local Universe. A simple model shows that the stellar mass content of GCLASS can evolve in typical distributions observed at lower redshifts if the clusters primarily accrete stellar mass onto the outskirts.}
        
   \maketitle
%

\hyphenation{in-tra-clus-ter}
\hyphenation{rank-or-der}

\section{Introduction}
One of the main objectives in the field of extragalactic astronomy is to understand the connection between galaxies and the distribution of the underlying dark matter. The growth of dark matter structures has been studied in large N-body simulations \citep[e.g.][]{springel05,boylankolchin09,navarro10}. From these simulations, the density profiles of collapsed structures have been found to be well represented by NFW-profiles \citep{NFW}. These profiles are described by two parameters: the halo mass, and the halo concentration parameter. The dependence of the concentration parameter $c$ on the halo mass, formation time and redshift has been studied with N-body simulations \citep[e.g.][]{wechsler02,neto07,duffy08,gao08}. These have shown that $c$, for the dark matter, is higher for lower mass haloes, higher for haloes that collapse early, and higher for haloes at lower redshift.

How baryons affect the distribution of the dark matter is still under debate \citep{dolag09,vandaalen11,newman13-2}. Baryons in the gas phase can cool and form stars at the bottom of the potential wells in the dark matter (sub-)haloes. The efficiency with which this happens depends on the properties of the halo \citep[see e.g.][]{kravtsov12,planelles13}. To constrain the physics behind these processes there is a number of key observables that can be exploited. In this paper we will concentrate on three of these, which we introduce in turn below, and will measure for a sample of 10 cluster sized haloes at $z \sim 1$.

First, to constrain the build up of stellar mass in central galaxies, we measure the stellar mass present in the central galaxies of GCLASS and compare it to direct measurements of their total halo masses. \citet{behroozi13} estimated the stellar mass in central galaxies versus total halo mass over a range of redshifts and halo masses in a \textit{statistical} way using the abundance matching technique. In this technique observables such as the stellar mass function and cosmic star formation history are combined with merger trees from dark matter simulations to provide constraints on the processes that build up the stellar mass in central galaxies. The stellar content of central galaxies, or Brightest Cluster Galaxies (BCGs) in the case of clusters, can grow by star-formation in the galaxy itself or by merging with other galaxies. Given the significant growth of stellar mass in BCGs as a function of redshift \citep{lin04bcg,lidman12}, this build-up is likely to occur through mainly dry mergers. However, the mass assembly has been shown \citep{lidman12} to increase more slowly than is expected from semi-analytic models \citep{delucia07}, but in good agreement with more recent simulations \citep{laporte13}. Since the main halo also accretes matter while the central galaxy is building up its stellar content, studies have focussed on the relationship between those processes. The \citet{behroozi13} estimates at $z=1.0$ cover a range of halo masses from $10^{11.3}<M_{h}/\rm{M_{\odot}}<10^{14.2}$, and are consistent with predictions from other abundance matching analyses \citep[e.g.][]{moster10,moster13}. In general the highest central stellar mass fraction is found in haloes of around $10^{12}\,\rm{M_{\odot}}$. By combining \textit{direct} measurements of total mass and stellar mass in the same haloes, we will test the results from abundance matching studies at $z\sim 1$. 

Second, a key measurement to understand the interplay between the growth of large scale structure and the formation and accretion of galaxies is to compare the total stellar mass as a function of halo mass. For a sample of groups selected at $0.1<z<1.0$ from COSMOS, \citet{giodini09} showed that their stellar mass fraction is a decreasing function of halo mass. Similar results are found by \citet{gonzalez07,gonzalez13}, \citet{andreon10} and \citet{hilton13} for samples of clusters around $z=0.1$, $z<0.1$ and $z=0.5$, respectively. Given that the most massive haloes are expected to grow by accreting lower mass systems, which have a higher stellar mass fraction, one would naively expect the stellar mass fraction of massive haloes to grow with cosmic time, even in the absence of in situ star-formation processes. Consequently, measurements on the stellar mass fraction in these haloes are used to constrain the progenitor population that form the building blocks of these haloes \citep{balogh08,mcgee09}. Due to the major caveats in comparing measurements from different studies with inhomogeneous data and different analyses, the relation itself is hard to constrain observationally \citep{leauthaud12,budzynski13}. So far little evolution with redshift has been found \citep{giodini09,lin12}. 

Third, the spatial distribution of the stellar mass component of satellites within the main halo is intimately related to accretion processes, and eventually the growth of the central galaxy. While the sub-haloes in pure gravitational N-body simulations get destroyed by tidal disruptions, the galaxies that have formed inside of them are more resistive to those forces \citep[e.g.][]{budzynski12}. The NIR luminosity and number density profiles are found to be described by NFW profiles for group-sized haloes \citep[e.g.][]{giodini09,tal13}, and clusters \citep[e.g.][]{carlberg97,lin04,muzzin07}. \citet{budzynski12} measured the radial distribution of galaxies from the SDSS around Luminous Red Galaxies in a redshift range $0.15<z<0.4$, and found that this distribution is also well described by an NFW profile. However, they found that the concentration parameter $c$ is lower for the galaxies than for the underlying dark matter. They found that the concentration is independent of mass, but that there is a hint of a mild dependence of the stellar mass concentration on redshift. 
A comparison of the radial stellar mass density distribution of clusters over a range of redshifts, linking high-$z$ systems to their likely descendants, yields insights in the evolution of the galaxy distribution. In this study we will extend the redshift baseline of these comparisons towards $z=1$.

We perform the aforementioned key measurements in an unexplored combination of redshift and halo mass range using the GCLASS survey, which contains deep 11-band photometry and spectroscopy for 10 rich clusters at $0.86<z<1.34$. This paper builds further on the results presented in several papers on the GCLASS sample. \citet[][hereafter M12]{muzzin12} present the spectroscopic sample, which is critical in this study to correct the photometrically selected galaxies by cluster membership. \citet[][hereafter vdB13]{vdB13} measure the stellar masses of the galaxies in the sample and present their stellar mass function (SMF). We will use the stellar masses estimated in this work for the current study. 
\citet{lidman12} identifies and studies the BCGs of GCLASS clusters as part of their analysis on the central galaxy stellar mass growth. 
The total GCLASS halo masses are estimated based on the velocity dispersions estimated in Wilson et al. (in prep.). To describe the masses of the clusters, we will use $R_{200}$ and $M_{200}$, which are defined as the radius at which the mean interior density is 200 times the critical density of the Universe, and the mass enclosed within this radius, respectively.

The structure of this paper is as follows. In Sect.~\ref{sec:sampledata} we present the GCLASS cluster sample, the available photometric and spectroscopic data, and give the results from a dynamical analysis to estimate the total halo masses. We also show how we obtain photometric redshifts and stellar mass estimates by summarizing the analysis from vdB13. We further show how the spectroscopic data are used to correct the full photometric catalogue for cluster membership. In Sect.~\ref{sec:bcgmasshalomass} we compare the stellar mass in the central galaxies with their halo masses. In Sect.~\ref{sec:results1} we present results on the total stellar mass versus halo mass relation of the clusters. In Sect.~\ref{sec:results2} we show how the galaxies are distributed radially and compare this to the expected dark matter profiles for these systems. Further, we discuss a possible evolutionary model to connect the $z\sim 1$ measurements to their likely descendants at lower redshift. In each section we compare the results with the literature and discuss how they are affected by possible systematics. We summarise and conclude in Sect.~\ref{sec:conclusions}.

All magnitudes we quote are in the AB magnitudes system and we adopt $\Lambda$CDM cosmology with $\rm{\Omega_m=0.3}$, $\rm{\Omega_{\Lambda}=0.7}$ and $\rm{H_0=70\, km\, s^{-1}\,  Mpc^{-1}}$. For stellar mass estimates we assume the Initial Mass Function (IMF) from \citet{chabrier03}.

\section{GCLASS Data \& Analysis}\label{sec:sampledata}\label{sec:analysis}
The GCLASS cluster sample consists of 10 rich clusters in the redshift range $0.86<z<1.34$ selected with the red-sequence selection method \citep{gladdersyee00} using the $z'-3.6\mu$m colour from the 42 square degree SpARCS survey \citep{muzzin09,wilson09}. These 10 clusters, which are amongst the richest at $z\sim 1$ in this survey area, are described in M12, and can be considered as a fair representation of IR-selected rich clusters within this redshift range. It is always a question how representative a cluster sample is of the full distribution of massive haloes, as it is impossible to select a sample based on halo mass. Each selection method has potential biases, whether it is X-ray selected, SZ-selected or galaxy-selected. However, especially at the high-mass end of the distribution, these selection methods are unlikely to cause significant biases in favour of particular types of galaxy clusters. Specifically, as e.g. \citet{blakeslee03} and \citet{mullis05} show, X-ray and SZ-selected clusters also show significant over-densities of red-sequence galaxies. We will discuss a possible selection bias further in Sect.~\ref{sec:discussion}. An overview of the GCLASS sample is given in Table~\ref{tab:overview}.

\begin{table*}[bt]
\caption{The 10 GCLASS clusters selected from SpARCS that form the basis of this study, with their dynamical properties.}
\label{tab:overview}
\begin{center}
\begin{tabular}{l l r r l l l c}
\hline
\hline
Name$^{\mathrm{a}}$ & $z_{\mathrm{spec}}$ &RA$^{\mathrm{b}}$&DEC$^{\mathrm{b}}$& $\sigma_{v}^{\mathrm{c}}$&$M_{200}^{\mathrm{d}}$& $R_{200}^{\mathrm{d}}$ & Spec-$z$\\
&&J2000&J2000&$[\rm{km/s}]$&[$10^{14}\,\mathrm{M_{\odot}}$]&[Mpc]&Members\\
\hline
SpARCS-0034&   0.867&00:34:42.06&-43:07:53.41&$\,\,\, 700_{- 150}^{+  90}$&$\,\,\, 2.4_{- 1.2}^{+ 1.0}$&$ 0.9_{- 0.2}^{+ 0.1}$& 45\\
SpARCS-0035&   1.335&00:35:49.70&-43:12:24.20&$\,\,\, 780_{- 120}^{+  80}$&$\,\,\, 2.5_{- 1.0}^{+ 0.9}$&$ 0.8_{- 0.1}^{+ 0.1}$& 20\\
SpARCS-0036&   0.869&00:36:45.03&-44:10:49.91&$\,\,\, 750_{-  90}^{+  80}$&$\,\,\, 2.9_{- 0.9}^{+ 1.0}$&$ 1.0_{- 0.1}^{+ 0.1}$& 47\\
SpARCS-0215&   1.004&02:15:24.00&-03:43:32.15&$\,\,\, 640_{- 130}^{+ 120}$&$\,\,\, 1.7_{- 0.8}^{+ 1.1}$&$ 0.8_{- 0.2}^{+ 0.2}$& 48\\
SpARCS-1047&   0.956&10:47:33.43&57:41:13.30&$\,\,\, 660_{- 120}^{+  70}$&$\,\,\, 1.9_{- 0.9}^{+ 0.7}$&$ 0.8_{- 0.2}^{+ 0.1}$& 31\\
SpARCS-1051&   1.035&10:51:11.21&58:18:03.17&$\,\,\, 500_{- 100}^{+  40}$&$\,\,\, 0.8_{- 0.4}^{+ 0.2}$&$ 0.6_{- 0.1}^{+ 0.1}$& 34\\
SpARCS-1613&   0.871&16:13:14.63&56:49:29.95&$1350_{- 100}^{+ 100}$&$16.9_{- 3.5}^{+ 4.0}$&$ 1.8_{- 0.1}^{+ 0.1}$& 92\\
SpARCS-1616&   1.156&16:16:41.32&55:45:12.44&$\,\,\, 680_{- 110}^{+  80}$&$\,\,\, 1.9_{- 0.8}^{+ 0.7}$&$ 0.8_{- 0.1}^{+ 0.1}$& 46\\
SpARCS-1634&   1.177&16:34:38.22&40:20:58.36&$\,\,\, 790_{- 110}^{+  60}$&$\,\,\, 2.9_{- 1.0}^{+ 0.7}$&$ 0.9_{- 0.1}^{+ 0.1}$& 50\\
SpARCS-1638&   1.196&16:38:51.64&40:38:42.91&$\,\,\, 480_{- 100}^{+  50}$&$\,\,\, 0.6_{- 0.3}^{+ 0.2}$&$ 0.5_{- 0.1}^{+ 0.1}$& 44\\
\hline
\end{tabular}
\end{center}
\begin{list}{}{}
\item[$^{\mathrm{a}}$] For full names we refer to \citet{muzzin12}.
\item[$^{\mathrm{b}}$] Coordinates of the BCGs, as identified by \citet{lidman12}.
\item[$^{\mathrm{c}}$] Velocity dispersions estimated by Wilson et al., in prep.
\item[$^{\mathrm{d}}$] Dynamical properties estimated using the relation between $\sigma_{v}$ and $M_{200}$ from \citet{evrard08}.
\end{list}
\end{table*}

The BCGs of these clusters have been identified and studied in \citet{lidman12}. In general the identification of the BCGs is straightforward, being the brightest cluster member in the $\rm{K_s}$-band, and we will use the same identification as done in \citet{lidman12}. In the cases of SpARCS-1051 and SpARCS-1634, \citet{lidman12} found that the BCGs are off-set from the approximate cluster centre by about 250kpc (projected).

The photometric data set consists of imaging in $ugriz\rm{JK_s}$ and 4 IRAC channels for each cluster. For details on the data reduction, and a description of the catalogue, we refer to vdB13. In summary, the catalogues contain objects detected in the $\rm{K_s}$-band, with Gaussian-weighted aperture fluxes in 11 filters to constrain the SEDs of the objects, and to separate stars from galaxies by combining their $u-\rm{J}$ and $\rm{J-K}$ colours. The depth of the images, and therefore the completeness of the catalogues, differs slightly from cluster to cluster. The median photometric completeness limit (80\%), in terms of stellar mass, is $10^{10.16}\,\rm{M_{\odot}}$ for the 10 clusters in the GCLASS sample.

Each cluster has substantial spectroscopic coverage provided by the GMOS instruments on Gemini North and Gemini South. The targets for spectroscopic follow-up were prioritized by their $3.6\mu$m flux and their projected cluster-centric distance, as explained in M12. The membership of the massive galaxies that constitute most of the stellar mass in the clusters are thus confirmed spectroscopically. Since the targeting completeness is well understood, we can use the sub-sample for which we have spectra to statistically correct the full catalogue for cluster membership. How this is done is outlined in vdB13 (Sect.~3.4), and expanded on in Sect.~\ref{sec:membershipcorr}.

\subsection{Total halo masses}\label{sec:halomasses}
\begin{figure}
\resizebox{\hsize}{!}{\includegraphics{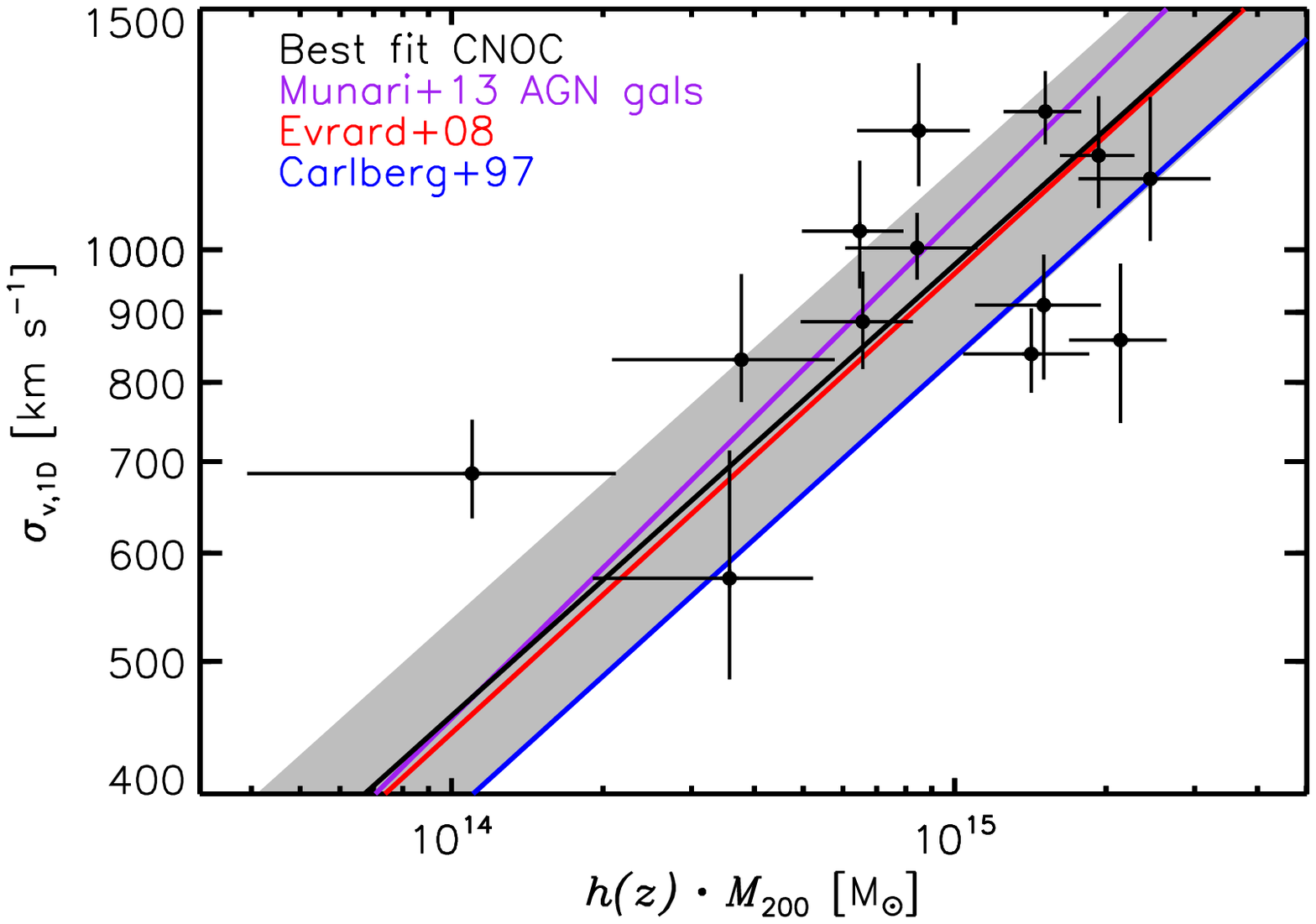}}
\caption{Measured velocity dispersion versus halo mass ($h(z) \cdot M_{200}$). Data points are measurements on the CNOC sample. Lensing masses are from \citet{hoekstra12} (which are revised in hoekstra et al., in prep.), whereas velocity dispersions are obtained from \citet{borgani99} and \citet{girardi01}. Although there is a substantial amount of intrinsic scatter (grey region indicates $\pm$1-$\sigma$ intrinsic scatter around the relation), the best fit to these data (black line) is very close to the \citet{evrard08} scaling relation (red).}
\label{fig:sigmavslensing}
\end{figure}

Using the sample of spectroscopically identified cluster galaxies, totalling 457 members for 10 clusters, we perform a dynamical analysis to estimate masses for each cluster. 
From the line-of-sight velocity distributions, which show approximately Gaussian profiles, the velocity dispersions are measured (Wilson et al., in prep.) using standard methods such as the shifting gapper and the bi-weight estimator \citep{beers90,girardi93,fadda96}, see Table~\ref{tab:overview}. Since we do not measure the velocity dispersion from dark matter particles but from subhaloes (or galaxies), several dynamical effects render this an imperfect tracer of the gravitational potential \citep[e.g.][]{saro13}. In an attempt to take account of these biases (which also depend on the spectroscopic target selection), various scalings between the velocity dispersion and halo mass ($M_{200}$) have been proposed in the literature \citep[e.g.][]{carlberg97b,evrard08,munari13}. These are of the form 
\begin{equation}
\sigma_{1\mathrm{D}} =\mathrm{A_{1D}}\left[\frac{h(z)\,M_{200}}{10^{15}\,\rm{M}_{\odot}}   \right]^{\alpha}\, \mathrm{km \, s^{-1}}, 
\end{equation}
where $\mathrm{A_{1D}}$ and $\alpha$ are parameters that are different for each study (Fig.~\ref{fig:sigmavslensing}).

In order to determine which scaling relation gives the best halo mass estimate for the measured velocity dispersions in GCLASS, we consider a sample of clusters which were originally studied as part of the Canadian Network for Observational Cosmology \citep[CNOC,][]{yee96}. A weak-lensing study has been performed for these systems, which provides for independent mass estimates \citep[][revised in Hoekstra et al., in prep.]{hoekstra07,hoekstra12}. For 13 of the clusters in this sample, velocity dispersions have been measured from spectroscopic targets that were chosen in a similar way as the targets selected in the GCLASS sample \citep{borgani99,girardi01}. Fig.~\ref{fig:sigmavslensing} compares the weak-lensing masses ($M_{200}$) to the line-of-sight cluster velocity dispersions. We fit a linear relation in this logarithmic plane, while fixing the slope to $\alpha=\frac{1}{3}$, and allow for the presence of intrinsic scatter in the fit. The black line shows the best-fitting scaling relation to the data points ($\mathrm{A_{1D}=972^{+60}_{-52}\,km\,s^{-1}}$), and we find a significant amount of intrinsic scatter around this relation ($\log(\sigma_{\sigma_{v}|h(z)\cdot M_{200}})=0.07^{+0.03}_{-0.02}$ dex).
The best-fitting scaling relation is very similar to the relation suggested by \citet{evrard08}. To estimate halo masses of the GCLASS clusters, we will therefore use the \citet{evrard08} scaling relation. This relation was also used by a recent dynamical study on the ACT cluster sample \citep{sifon13}, which simplifies a comparison with the results from this sample \citep[e.g.][]{hilton13} in the rest of this paper. Values of $M_{200}$ and $R_{200}$ are shown in Table~\ref{tab:overview}. Statistical uncertainties are given (propagated from uncertainties on the velocity dispersion), but note that there is also a significant systematic uncertainty ($\sim 20\%$), corresponding to the choice of scaling relation, and indicated by the substantial amount of intrinsic scatter. Note that the $R_{200}$ values have a smaller fractional uncertainty, since $R_{200}\propto M_{200}^{1/3}$.



\subsection{Photometric redshifts and Stellar masses}\label{sec:stellarmasses}
We estimate photometric redshifts for all galaxies in the $\rm{K_s}$-band selected catalogue using the EAZY code \citep{brammer08}. In vdB13 we assessed the performance by comparing the photo-$z$ estimates to spec-$z$ measurements for the galaxies that have been observed spectroscopically. We found a scatter of $\sigma_z=0.036$ in $\frac{\Delta z}{1+z}$, a negligible bias and fewer than 5\% outliers.  

After fixing the redshift for each object at its spec-$z$, or the photo-$z$ when a spec-$z$ is not available, we estimate stellar masses using FAST \citep{kriek09}. The stellar population libraries from \citet{bc03} are used to obtain the model SED that gives the best fit to the photometric data. We use a parameterization of the star formation history as $SFR \propto e^{-t/\tau}$, where the time-scale $\tau$ is allowed to range between 10 Myr and 10 Gyr. We also assume a \citet{chabrier03} IMF, solar metallicity, and the \citet{calzetti00} dust law. For estimates on the stellar-mass completeness of the catalogues, we refer to vdB13. To approximate the statistical uncertainty on each stellar mass measurement, we perform 100 Monte-Carlo simulations in which we perturb the photometric aperture flux measurements within their estimated errors. Each realisation of the catalogue gives a slightly different SED fit, and therefore the mass-to-light ratio ($M_{\star}/L$) is different. We translate the obtained scatter in $M_{\star}/L$ into an approximate uncertainty on the stellar mass, after including uncertainties on the spectral templates themselves. We find typical uncertainties on $M_{\star}/L$ of 0.21 dex at $M_{\star}\sim 10^{10}\,\rm{M_{\odot}}$, and 0.13 dex at $M_{\star}\sim 10^{11}\,\rm{M_{\odot}}$.

\subsection{Cluster membership correction}\label{sec:membershipcorr}

\begin{figure}
\resizebox{\hsize}{!}{\includegraphics{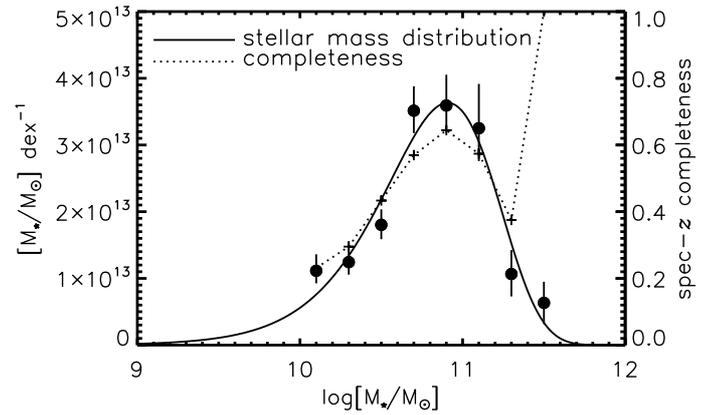}}
\caption{Solid line: the distribution of total stellar mass contained in galaxies with a given stellar mass. The points with error bars are the measurements of the SMF presented in \citet{vdB13}, but integrated over the mass bins. Dotted line: the spectroscopic completeness for galaxies with projected distances from the BCG less than $R_{200}$. For the galaxies that constitute most of the stellar mass in the clusters, the spectroscopic completeness is high ($\gtrsim 50\%$).}
\label{fig:stelmassdist}
\end{figure}

Fig.~\ref{fig:stelmassdist} shows the distribution of total stellar mass contained in galaxies with a given stellar mass (solid line). This line is based on the best-fitting Schechter function for the total galaxy population from vdB13. The points with error bars are the measurements of the SMF presented there, and are integrated masses over the SMF in each bin, i.e. $\int_{M_{\rm{min}}}^{M_{\rm{max}}} \Phi(M) \cdot dM$, where $\Phi(M)$ represents the number density of galaxies as a function of stellar mass. With the characteristic mass of the Schechter function around $M_{\star}=10^{11}\,\rm{M_{\odot}}$, galaxies around this mass contribute most to the total stellar mass of the cluster. The dotted line shows the spectroscopic completeness for galaxies with projected distances from the BCG less than $R_{200}$, and shows that for the galaxies with stellar masses around $M_{\star}=10^{11}\,\rm{M_{\odot}}$, the completeness is high ($\ge 50\%$). For measurements within $R_{500}$ the completeness is even higher. For that reason, the measurements of the total stellar mass of the clusters are based mostly on spectroscopic redshifts, and are robust with respect to how we correct the photometric sample for completeness.

We use the limited number of galaxies in the fields that have been targeted spectroscopically to estimate the probability that a galaxy is part of the cluster for the objects that do not have a measured spectroscopic redshift. For objects with stellar masses exceeding $\sim 10^{10}\,\rm{M_{\odot}}$ that were targeted, the success rate of obtaining a reliable spec-$z$ is higher than 90\% (M12). Given that the targeting prioritization is known (M12), we can correct the photometrically selected sample for cluster membership using the sub-sample of spectroscopic targets. To do this we take a similar approach as outlined in vdB13 (Sect.~3.4), but with a few adaptations. 

\begin{figure}
\resizebox{\hsize}{!}{\includegraphics{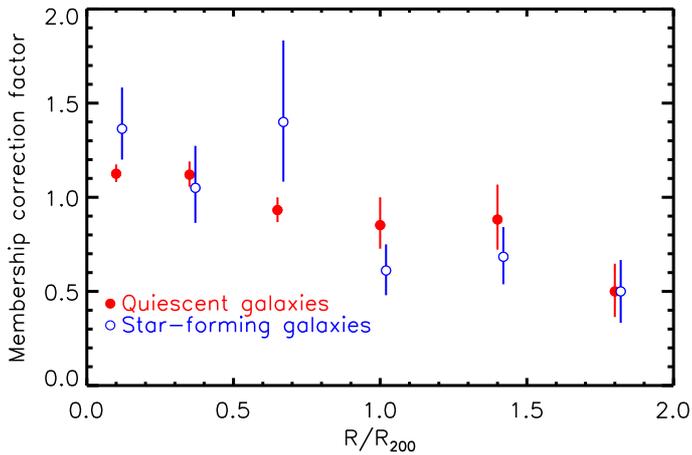}}
\caption{Correction factors as a function of radius, scaled by $R_{200}$, for the cluster ensemble. Error bars represent uncertainties estimated from Monte-Carlo simulations. Further away from the projected centres, the correction factors go down because galaxies are increasingly more likely to be part of the field.}
\label{fig:corrfactors_radius_r200}
\end{figure}

The radial distance of each galaxy is rescaled to units of $R_{200}$, instead of physical distance. Then we measure for the cluster ensemble, in bins of radial distance and stellar mass, the fraction of correctly identified cluster galaxies based on their photo-$z$. Comparing this number to the total number of spec-$z$ selected cluster members in this bin, we obtain membership correction factors that are used to correct the photometrically selected numbers for membership. The correction factors as a function of radial distance are shown in Fig.~\ref{fig:corrfactors_radius_r200}. The membership correction factors are a decreasing function of distance, since the clusters are less overdense further away from their cluster centre. The blue (red) points represent the population of star-forming (quiescent) galaxies. 
For the correction factors as a function of stellar mass we refer to vdB13 (Fig.~4).

To further improve the estimates on the total stellar mass associated with each cluster, we estimate the contamination by field galaxies for each individual cluster. This minor correction to the photo-$z$ selected sample for each cluster is due to cosmic variance, slight differences in photometric redshift quality between the fields, and also the dependence of angular size associated with $R_{200}$ on the cluster mass and redshift. To estimate this overdensity parameter we 1) apply the correction factors that we use on the photometric sample (e.g. Fig.~\ref{fig:corrfactors_radius_r200} and vdB13 (Fig.~4)) on all spectroscopically targetted galaxies, then 2) use this to estimate the number of cluster members in this sample, and 3) divide the actual number of spectroscopic cluster members by the estimated number of cluster members to give the correction factor. This cluster overdensity parameter is by construction around 1.0 and ranges from 0.86 to 1.22 for the clusters in our sample.

\section{Central stellar mass versus halo mass}\label{sec:bcgmasshalomass}
\begin{figure}
\resizebox{\hsize}{!}{\includegraphics{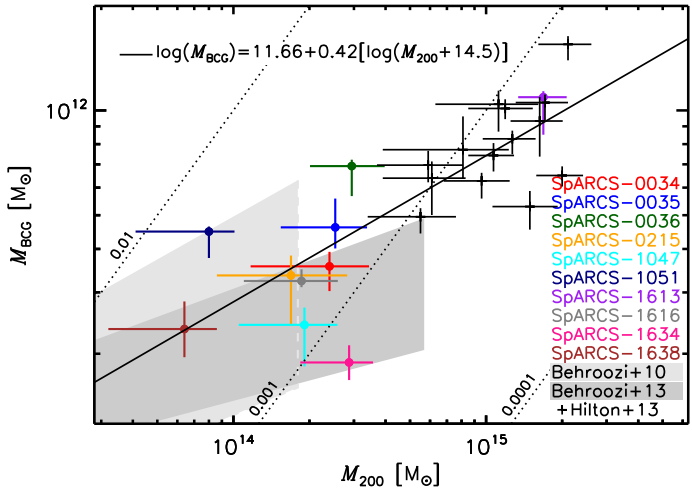}}
\caption{BCG stellar mass versus total halo mass. Black dots show lines of constant stellar mass fractions of 0.0001, 0.001 and 0.01. $^\backprime+^\prime$-signs show the results from \citet{hilton13}. The relation is the best fit to the combined data set of \citet{hilton13} and the current study. Estimates from \citet{behroozi10} and \citet{behroozi13} are indicated by the shaded regions.}
\label{fig:BCGvshalomass_galfitcorr}
\end{figure}

In Fig.~\ref{fig:BCGvshalomass_galfitcorr} we show the stellar mass of the central galaxy as a function of dynamical halo mass. Stellar masses are measured based on $M_{\star}/L$'s estimated with FAST, multiplied with the total flux in the $\rm{K_s}$-band. Since brightest cluster galaxies (BCGs) generally have extended light profiles, their flux measured with {\tt SExtractor} in Kron elliptical apertures is under-estimated. To account for the total flux of the BCGs in the $\rm{K_s}$-band, we use {\tt GALFIT} to fit S\'ersic profiles to these galaxies. We make sure that we carefully mask any nearby satellite galaxies and perform 10 different fits where we convolve the profiles with different stars to approximate uncertainties due to the PSF. We compare the integrated flux in these S\'ersic profiles with the {\tt SExtractor} magnitudes in Table~\ref{tab:galfit}. The values show the median values and the maximum and minimum values for the 10 different {\tt GALFIT} runs, after rejecting the highest and lowest value. The difference between the {\tt GALFIT} and {\tt SExtractor} measurements is typically about 0.2 mag, and depends mainly on the shape of the profile, which is described by the S\'ersic parameter $n$. To obtain the total stellar masses of the central galaxies we multiply the total flux in the $\rm{K_s}$-band with the $M_{\star}/L$ estimated using FAST, and include both the flux-error and the error on $M_{\star}/L$ (which is the dominant source of uncertainty).

\begin{table}[h]
\caption{$\rm{K_s}$-band magnitudes for the BCGs identified in \citet{lidman12} for the GCLASS clusters. The last column gives the stellar masses corresponding to the {\tt GALFIT} total integrated magnitude, and the errors also include the statistical uncertainty on $M_{\star}/L$.}
\label{tab:galfit}
\begin{center}
\begin{tabular}{l l l l l}
\hline
\hline
Name & $\rm{MAG\_AUTO}$ & \tt{GALFIT} &\tt{GALFIT}&$M_{\star,BCG}$\\
&$[\rm{mag_{AB}}]$&$[\rm{mag_{AB}}]$&S\'ersic - $n$&$[10^{11}\,\rm{M_{\odot}}]$\\
\hline
SpARCS-0034&$   16.59\pm    0.01$&$   16.51_{-    0.03}^{+    0.04}$&$  3.68_{-  0.37}^{+  0.36}$&$\,\,\,  3.56_{-  0.54}^{+  0.36}$         \\
SpARCS-0035&$   17.27\pm    0.01$&$   17.06_{-    0.01}^{+    0.01}$&$  3.77_{-  0.13}^{+  0.13}$&$\,\,\,  4.61_{-  0.60}^{+  0.97}$         \\
SpARCS-0036&$   16.40\pm    0.01$&$   16.10_{-    0.04}^{+    0.01}$&$  3.82_{-  0.37}^{+  0.30}$&$\,\,\,  6.92_{-  1.25}^{+  0.30}$         \\
SpARCS-0215&$   17.05\pm    0.01$&$   16.86_{-    0.02}^{+    0.02}$&$  3.02_{-  0.13}^{+  0.14}$&$\,\,\,  3.36_{-  0.94}^{+  0.47}$         \\
SpARCS-1047&$   17.29\pm    0.01$&$   17.03_{-    0.03}^{+    0.01}$&$  4.35_{-  0.41}^{+  0.22}$&$\,\,\,  2.42_{-  0.58}^{+  0.29}$         \\
SpARCS-1051&$   17.11\pm    0.02$&$   16.73_{-    0.04}^{+    0.03}$&$  6.87_{-  0.93}^{+  0.97}$&$\,\,\,  4.49_{-  0.72}^{+  0.15}$         \\
SpARCS-1613&$   15.67\pm    0.01$&$   15.50_{-    0.01}^{+    0.00}$&$  3.25_{-  0.15}^{+  0.10}$&$ 10.91_{-  2.40}^{+  0.44}$         \\
SpARCS-1616&$   17.01\pm    0.01$&$   16.96_{-    0.02}^{+    0.01}$&$  3.03_{-  0.15}^{+  0.18}$&$\,\,\,  3.24_{-  0.13}^{+  0.26}$         \\
SpARCS-1634&$   17.41\pm    0.01$&$   17.42_{-    0.01}^{+    0.01}$&$  0.83_{-  0.01}^{+  0.01}$&$\,\,\,  1.89_{-  0.21}^{+  0.23}$         \\
SpARCS-1638&$   17.71\pm    0.02$&$   17.43_{-    0.01}^{+    0.01}$&$  5.23_{-  0.12}^{+  0.12}$&$\,\,\,  2.36_{-  0.40}^{+  0.47}$         \\
\hline
\end{tabular}
\end{center}
\end{table}

Considering the GCLASS data in Fig.~\ref{fig:BCGvshalomass_galfitcorr}, we find mild evidence for a correlation between the BCG stellar mass and halo mass, with a Spearman rank coefficient $\rho =0.49$. The fraction of mass contained in stellar form in the BCG is approximately 0.001 of the halo mass. 

\citet{behroozi10} and \citet{behroozi13} estimated the stellar mass versus virial halo mass relation over a range of redshifts and halo masses using the abundance matching technique. At the high mass end we make a comparison between their estimates and our observations, which are based on direct measurements of the total halo masses and stellar masses of centrals in the same systems. We multiply the Behroozi halo masses by factor 1.11 to account for the difference between their virial halo masses and $M_{200}$ \citep{bryan98}. We show the \citet{behroozi10} (\citet{behroozi13}) prediction for $z=1$ by the light (dark) shaded area in Fig.~\ref{fig:BCGvshalomass_galfitcorr}. Although the allowed areas are large due to inclusions of statistical and systematic uncertainties, the results from \citet{behroozi10} seem to be in better agreement with the GCLASS data than the results from \citet{behroozi13}. What is different in both abudance matching studies is the specific treatment of intra-cluster light (ICL) in \citet{behroozi13}. When a galaxy merger occurs in this new model, the stars associated with the satellite galaxy may either be deposited onto the central galaxy, or be ejected into the ICL. Since \citet{behroozi13} estimate the ICL to be of a significant contribution to the total stellar mass at $z=1$, this is potentially related to an under-prediction of the stellar mass in the central galaxies.

To increase the dynamic range in terms of cluster halo mass, in order to constrain the power-law slope of this relation, we compare our results to those from \citet{hilton13}, which were obtained from a sample of ACT SZ-selected clusters. To be able to compare the results directly, we reduce the stellar masses estimated from \citet{hilton13} by 0.24 dex to account for differences in the adopted IMF. Note that \citet{hilton13} did not fit the SED of the BCG with a model to constrain $M_{\star}/L$, but rather assumed a single burst stellar population that has a formation redshift $z_{\mathrm{f}}=3$. For the purpose of estimating BCG stellar masses the difference between these approaches is small ($<$0.1 dex), because the BCGs contain relatively old stellar populations. The $M_{200}$ measurements for this cluster sample are taken from \citet{sifon13}. 
Fig.~\ref{fig:BCGvshalomass_galfitcorr} shows a clear relation between the BCG stellar mass and total halo mass from GCLASS and \citet{hilton13}.

When we fit a slope to the combined set of data points, we have to account for intrinsic scatter in the relation to ensure that we do not give too much weight to precise measurements that are far off the mean relation. We follow the approach outlined in \citet{hoekstra11} to perform a three parameter fit to these data points. Besides the parameters describing the power-law relation, the intrinsic scatter is assumed to be described by a log-normal distribution, for which we fit the dispersion $\sigma$. The intrinsic scatter is best described by $\log(\sigma_{M_{BCG}|M_{200}})=0.12^{+0.03}_{-0.02}$ dex, and the best-fitting relation is $\log(M_{BCG})=(11.66\pm0.03)+0.42^{+0.06}_{-0.07}\cdot[\log(M_{200})-14.5]$. This relation is plotted in Fig.~\ref{fig:BCGvshalomass_galfitcorr} and indicates that the BCG stellar mass fraction is lower for higher mass haloes. The fit shows that there is a significant amount of intrinsic scatter in the relation between central galaxy stellar mass and halo mass, which is consistent with the finding of e.g. \citet{leauthaud12-1}.

Note that our data do not allow for measurements of the intracluster light (ICL), and therefore the contribution of intracluster stars to the central stellar mass is neglected. Formally the measured values are therefore lower limits, but \citet{burke12} show that the contribution of intracluster stars to the total stellar mass at $z\sim 1$ is expected to be significantly smaller than at lower redshifts. 
On the contrary, \citet{behroozi13} suggest a picture in which a significant fraction of the ICL has already been formed at $z\sim 1$. Note however the slight tension between their statistical study and our observations of the stellar mass in the central galaxy (which is related to the build-up of the ICL component) in Fig.~\ref{fig:BCGvshalomass_galfitcorr}.

\citet{lidman12} measure the BCG stellar mass versus halo mass for a sample of 160 BCGs in the redshift interval $0.03<z<1.63$. Besides the different redshift range they study, their analysis is slightly different from ours. \citet{lidman12} constrain the $M_{\star}/L$ of the BCGs with J-, and $\rm{K_{s}}$-band data and do not use {\tt GALFIT} to probe the extended light profiles of the BCGs. The slope fitted by \citet{lidman12} is $M_{200}\propto M_{\rm{BCG}}^{\,\,1.6\pm0.2}$. The reciprocal of this is consistent with our slope to within 2-$\sigma$. 

\section{Total stellar mass versus halo mass}\label{sec:results1}
We make a comparison between the halo masses and the total stellar mass in the halo, including the satellites. We will perform all measurements both within $R_{200}$ and within $R_{500}$ to provide a reference and facilitate the comparison with literature measurements. When necessary, we will convert between $R_{200}$ and $R_{500}$ by applying the concentration parameter estimated from \citet{duffy08}. For the mass and redshift range of the GCLASS clusters, \citet{duffy08} find a typical concentration of $c=2.7$, which is consistent with a stacked weak-lensing measurement of clusters at $z\sim 1$ \citep{sereno13}. Corresponding to this concentration parameter, we will use the relationships $R_{500}=0.632\cdot R_{200}$ and $M_{500}=0.631\cdot M_{200}$. 

For each cluster we sum the stellar mass contained in galaxies with a spectroscopic redshift consistent with the cluster that exceed the mass completeness limit of the cluster. The $\rm{K_{s}}$-band flux limits were simulated for each cluster, and corresponding stellar mass completeness limits were estimated and presented in vdB13 (Table~1). To this we add the photo-$z$ selected sources that we correct for cluster membership using the method explained in Sect.~\ref{sec:membershipcorr}, provided that their projected radii from the BCG are less than $R_{200}$ (or $R_{500}$). Since the overdensity of the cluster with respect to the field is different for each cluster, as explained in Sect.~\ref{sec:membershipcorr}, we correct the total stellar mass of the photometric sample with the cluster overdensity parameter for each cluster.

The stellar mass is now measured within a projected radius of $R_{200}$ (or $R_{500}$), but to estimate the stellar mass fraction and be able to compare to results in the literature we have to deproject the stellar mass onto a sphere with radius $r_{200}$ (or $r_{500}$), since the halo mass $M_{200}$ (or $M_{500}$) is defined in that way. Assuming a concentration parameter $c=2.7$ and integrating the NFW profile along the line of sight, we find that 74\% of the mass in the cylinder also lies within the sphere with radius $R_{200}$ (and 69\% when we make this comparison for $R_{500}$). We therefore multiply the stellar mass estimates by a factor 0.74 (0.69 for $R_{500}$).

Since so far we only considered galaxies with stellar masses exceeding the mass completeness limits, we have to estimate the stellar mass contained in lower mass galaxies. We measured the Schechter parameters of the SMF in vdB13, and although these parameters were constrained by galaxies with stellar masses exceeding $10^{10}\,\rm{M_{\odot}}$, we use the integral of this Schechter function for masses below the stellar mass completeness limits to correct for these lower mass galaxies. Fig.~\ref{fig:stelmassdist} shows that the total stellar mass contained in low-mass galaxies is small. The percentage by which we correct the stellar mass depends on the stellar mass completeness and ranges from $4\%$ for SpARCS-0035 to $25\%$ for SpARCS-0036. Given the size of these corrections factors, they do not have a significant effect on the results, especially because the depth in terms of stellar mass is independent of the redshift or halo mass of the clusters. Total stellar masses are listed in Table ~\ref{tab:totstelmass}.

\begin{figure*}
\resizebox{\hsize}{!}{\includegraphics{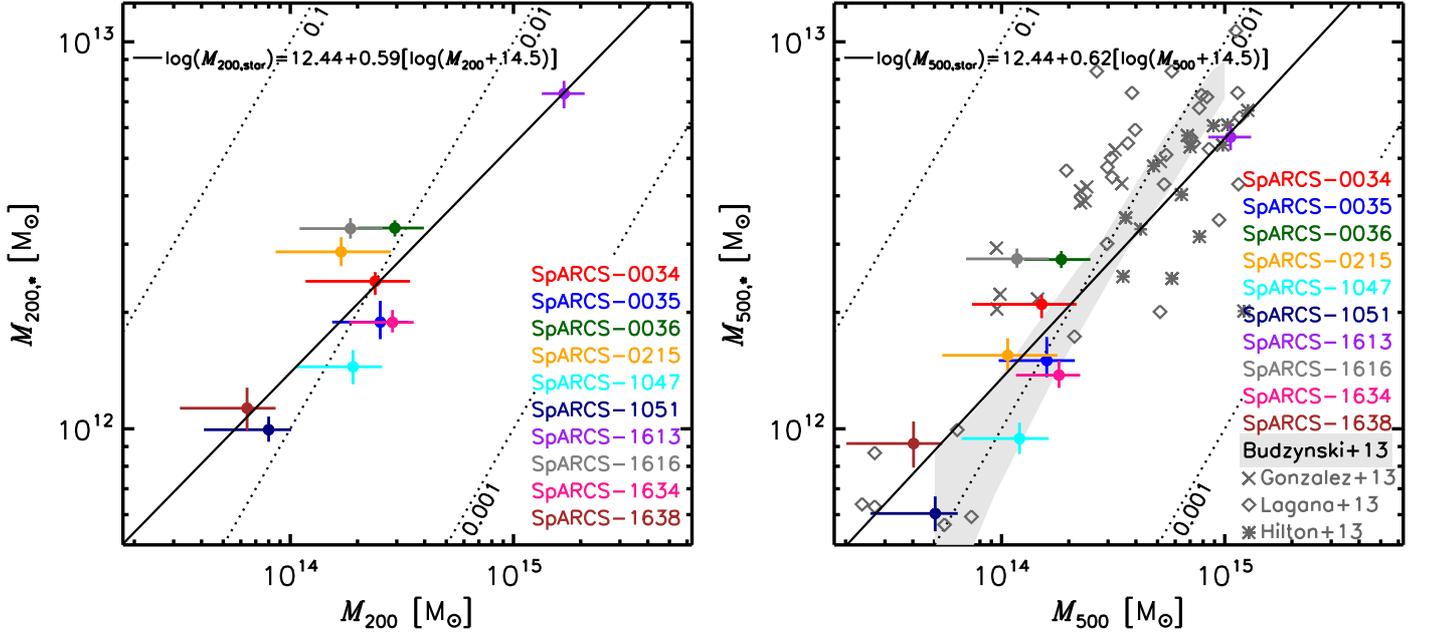}}
\caption{Total stellar mass versus halo mass within a sphere with radius $R_{200}$ and $R_{500}$ (left and right panels, respectively). Error bars represent uncertainties on individual mass measurements and uncertainties on the membership correction for galaxies we do not have spectra for. Dotted lines show locations with constant stellar mass fractions. The literature measurements (right panel) are measured over a range of redshifts, and are based on different analyses. When possible, the data points are corrected for differences in IMF and $M_{\star}/L$'s, as explained in the text.}
\label{fig:stellmassvshalomass}
\end{figure*}

In Fig.~\ref{fig:stellmassvshalomass} we show the total stellar mass versus total halo mass compared within $R_{200}$ and $R_{500}$ (left and right panels, respectively) for the GCLASS systems. Error bars in the vertical direction include statistical uncertainties on individual stellar mass measurements, and uncertainties on the estimated probabilities that a photometrically selected galaxy is part of the cluster. The latter uncertainty, which dominates, includes the error on the overdensity parameter for each cluster. The GCLASS data show a clear correlation, with spearman coefficient $\rho = 0.65$ (within $R_{200}$), and $\rho = 0.62$ (within $R_{500}$). 

We fit a power-law relation to the GCLASS data points, with the amount of intrinsic scatter as a free parameter, and described by a log-normal distribution with scatter $\sigma$. We find the following best-fitting parameters for the comparison within $R_{200}$; $\log(\sigma_{M_{200,\star}|M_{200}})=0.08^{+0.04}_{-0.05}$ dex, and the relation $\log(M_{200,\star})=(12.44\pm0.04)+(0.59\pm0.10)\cdot[\log(M_{200})-14.5]$. When we perform the fit to the data within $R_{500}$ we find; $\log(\sigma_{M_{500,\star}|M_{500}})=0.11^{+0.05}_{-0.04}$ dex, and the relation $\log(M_{500,\star})=(12.44^{+0.05}_{-0.06})+(0.62\pm0.12)\cdot[\log(M_{500})-14.5]$. Both relations are shown in Fig.~\ref{fig:stellmassvshalomass}. The slope of the relation is consistent with the slope found by \citet{lin12}, who measured it to be $0.71\pm0.04$ for a sample of redshift $z<0.6$ clusters. The small amount of intrinsic scatter in the relation between total stellar mass and halo mass indicates that stellar mass is a good proxy for total halo mass (albeit with large measurement uncertainties on individual clusters), as was also suggested by \citet{andreon12}. 

For 6 X-ray selected galaxy clusters at $z\sim 1$, \citet{burke12} show that the contribution of the ICL to the total J-band flux within $R_{500}$ is about 1-4\%. Since this contribution is much (factor $\sim$2-4) smaller than the contribution of the ICL at low-$z$, our measurements should be close to the actual mass in stars. 

Given that this tight relation between total stellar mass and halo mass already exists at $z\sim 1$, and that the stellar mass fraction is decreasing with increasing halo mass, one would naively expect the stellar mass fraction of these massive haloes to increase towards lower redshifts. That is because the likely systems that will be consumed by these haloes are those with a high stellar mass fraction \citep{mcgee09}. In this simple picture the stellar mass fraction would increase, even in the absence of in-situ star formation. Given this naive expectation, it is therefore interesting to make a comparison of the stellar mass content of haloes at lower redshifts.

\begin{table}[h]
\caption{Total stellar masses projected onto spheres with radii $R_{200}$ and $R_{500}$ for the GCLASS clusters.}
\label{tab:totstelmass}
\begin{center}
\begin{tabular}{l r r r}
\hline
\hline
Name & $M_{200,\star}$ & $M_{500,\star}$ &$M_{500,\star} (3.6\rm{\mu m})^{\mathrm{a}}$\\
&$[10^{12}\,\rm{M_{\odot}}]$&$[10^{12}\,\rm{M_{\odot}}]$&$[10^{12}\,\rm{M_{\odot}}]$\\
\hline
SpARCS-0034&$  2.40^{+  0.16}_{-  0.15}$&$  2.10^{+  0.14}_{-  0.14}$&-\\
SpARCS-0035&$  1.89^{+  0.22}_{-  0.20}$&$  1.50^{+  0.18}_{-  0.16}$&$  5.43^{+  2.92}_{-  1.90}$\\
SpARCS-0036&$  3.30^{+  0.16}_{-  0.15}$&$  2.74^{+  0.14}_{-  0.13}$&-\\
SpARCS-0215&$  2.86^{+  0.25}_{-  0.23}$&$  1.55^{+  0.15}_{-  0.14}$&-\\
SpARCS-1047&$  1.45^{+  0.15}_{-  0.13}$&$  0.94^{+  0.09}_{-  0.08}$&-\\
SpARCS-1051&$  1.00^{+  0.07}_{-  0.07}$&$  0.60^{+  0.06}_{-  0.06}$&-\\
SpARCS-1613&$  7.35^{+  0.60}_{-  0.55}$&$  5.68^{+  0.42}_{-  0.39}$&$ 18.72^{+  9.65}_{-  6.37}$\\
SpARCS-1616&$  3.29^{+  0.20}_{-  0.19}$&$  2.75^{+  0.16}_{-  0.15}$&$  7.14^{+  2.19}_{-  1.68}$\\
SpARCS-1634&$  1.88^{+  0.13}_{-  0.12}$&$  1.38^{+  0.11}_{-  0.10}$&$  3.37^{+  2.58}_{-  1.46}$\\
SpARCS-1638&$  1.13^{+  0.14}_{-  0.13}$&$  0.92^{+  0.13}_{-  0.11}$&$  2.33^{+  1.56}_{-  0.93}$\\
\hline
\end{tabular}
\end{center}
\begin{list}{}{}
\item[$^{\mathrm{a}}$] Taking the background subtracted flux in IRAC 3.6$\rm{\mu m}$ and assuming the same $M_{\star}/L$ for every galaxy in each cluster, based on a single burst stellar population with $\tau=0.1 \,\rm{Gyr}$ formed at $z_{\mathrm{f}}=3$. 
\end{list}
\end{table}

\subsection{Comparison to other samples}
We compare our measurements to others in the literature (mostly performed at low-$z$) for $R_{500}$, since this radius was used by most studies that estimate the halo masses with X-ray data. However, there are several important caveats to make before we can make a fair comparison. The adopted $M_{\star}/L$ is a major systematic uncertainty in any study and depends on the assumed IMF due to differences in the contribution of low mass stars to the total mass. We transform the results from other studies to the Chabrier IMF by subtracting 0.24 dex in mass for a Salpeter IMF, or adding 0.04 dex to the mass for a Kroupa IMF. The $M_{\star}/L$ depends on galaxy type, but due to the lack of multi-wavelength photometry, it is often assumed that all cluster galaxies are composed of the same stellar population. If one assumes an old stellar population (and therefore a high $M_{\star}/L$), the mass of the late-type galaxies (and thus the cluster as a whole) is over-estimated. Such an effect will be more pronounced at higher redshift because of the higher number density of late-type galaxies in high-$z$ clusters (M12, vdB13). We will point out possible issues for each of the comparison samples below.

An obvious study to compare our results to is based on an SZ-selected cluster sample from the ACT, with a redshift range overlapping with GCLASS and a median redshift of $z=0.50$ \citep{hilton13}. A complication is that \citet{hilton13} estimated cluster stellar masses based on the total IRAC 3.6$\rm{\mu m}$ flux measured after a statistical background subtraction. Instead of fitting a $M_{\star}/L$ for each galaxy based on SED modelling, they assume a stellar population that is formed at $z_{\mathrm{f}}=3$, following a $\tau=0.1 \,\rm{Gyr}$ single burst model and the \citet{bc03} stellar population synthesis model.  
To estimate the effects of these assumptions and see if this creates a bias, we follow the method described by \citet{hilton13} to obtain the background subtracted IRAC 3.6$\rm{\mu m}$ flux within $R_{500}$ for the 5 GCLASS clusters for which we have deep IRAC data (vdB13), and estimate the total stellar mass based on the described stellar population. Table~\ref{tab:totstelmass} compares these estimates with the total stellar mass in the clusters obtained by the full SED fitting analysis. The approach with a fixed $M_{\star}/L$ over-estimates the stellar mass in all clusters by at least a factor of 2, and this difference seems to be largest for the highest redshift cluster. This is consistent with the notion that the blue fraction, and therefore the fraction of galaxies with relatively low $M_{\star}/L$, increases with redshift \citep[cf.][]{boeffect78}. It is also possible that the stellar population assumed by \citet{hilton13} has a formation redshift ($z_{\mathrm{f}}=3$) that is too high. 
After correcting the stellar masses from \citet{hilton13} to a Chabrier IMF, we divide them by an additional factor of 2 as an approximate correction for the $M_{\star}/L$ explained above. These data points are overplotted in Fig.~\ref{fig:stellmassvshalomass} (right panel, $\divideontimes$-symbols), and lie around the relation that is the best fit to the GCLASS data. Note that since we used the red-sequence selected GCLASS sample to measure this bias, the real bias might be even larger if the SZ-selected sample contains a lower fraction of quiescent galaxies.

To study a possible evolution in the stellar mass content of clusters we consider \citet{lagana13}, who measure the stellar mass content in a sample of $z<0.3$ clusters. Estimates for $M_{500}$ are obtained from X-ray observations. To measure the total stellar mass from the available SDSS data, the galaxy population is separated between early-type and late-type galaxies using the ($u-i$) colour. Exploiting the $M_{\star}/L$ from \citet{kauffmann03} in the $i$-band for these galaxy types, \citet{lagana13} estimate stellar masses. Since the \citet{kauffmann03} $M_{\star}/L$'s are based on the Kroupa IMF, we subtract 0.04 dex to compare their results to ours, and overplot them in Fig.~\ref{fig:stellmassvshalomass} (right panel, $\Diamond$-symbols).

Another nearby cluster sample is the one studied by \citet{gonzalez07}, which is in the range $0.03<z<0.13$, and these measurements are revised in \citet{gonzalez13}. In these studies, a single $M_{\star}/L$ was used for each galaxy, irrespective of their type. From a dynamical analysis of the SAURON project, they estimate the average $M_{\star}/L$ in the $i$-band, which they found to be lower than the $M_{\star}/L$ based on an assumed Salpeter IMF. We correct their $M_{\star}/L$ to a Chabrier IMF by subtracting 0.12 dex, and overplot the points from \citet{gonzalez13} in Fig.~\ref{fig:stellmassvshalomass} (right panel, $\times$-symbols). The stellar mass fractions they find are in approximate agreement with the stellar mass fractions of the GCLASS clusters, although they find a somewhat shallower slope of $0.52\pm0.04$ when they fit a relation to only their data set. Given that the fraction of red (with a large $M_{\star}/L$) galaxies depends on halo mass, it is possible that this slope is biased due to the assumption of a single $M_{\star}/L$ for the sample.

To increase the dynamic range of the comparison samples, we make a comparison to the measurements from \citet{budzynski13}, who measured the stellar mass fraction across a wide range of masses in the group and cluster regime from the SDSS. Their stacked measurement of over 20,000 optically selected systems at $0.15<z<0.4$ is shown by the shaded region in Fig.~\ref{fig:stellmassvshalomass}. Since their analysis is very similar to our, we do not have to correct their measurements for differences in e.g. $M_{\star}/L$. Both the normalisation and their slope of $0.89\pm0.14$ are consistent with the relation we find for GCLASS. When they stack original SDSS images to measure the contribution from the ICL to the stellar mass in their sample, they find a slope that is even steeper. 

We note that there are caveats that arise when comparing different cluster samples, as was also pointed out by several other studies \citep[e.g.][]{leauthaud12,budzynski13}. Performing the analysis described by \citet{hilton13} on the GCLASS data shows that there is a bias in the total stellar mass when a single $M_{\star}/L$ is assumed for all cluster galaxies, especially at high-$z$. This bias in the stellar mass can be larger than the evolution expected in the redshift range $0<z<1$. This shows that it is important to analyse the full SED of each galaxy to estimate its stellar mass. Thanks to the spectroscopic coverage of the GCLASS sample, which is more than 50\% complete within $R_{200}$ for the galaxies that dominate the total stellar mass content, membership assignment is relatively straight-forward. In other analyses, where a statistical background subtraction is performed, this can be a major uncertainty for individual systems. We attempted to correct for differences in the analyses between literature studies to be able to compare the total stellar mass fractions between different epochs. Within the uncertainties there seems to be a good agreement between the studies over this redshift range, showing that there is no significant evolution in the stellar mass fraction at fixed halo mass in the redshift range $0<z<1$. To tighten the constraints on a possible evolution of this relation, a large and more homogeneous dataset and analysis are required.

\section{Radial stellar density distribution}\label{sec:results2}
Measurements of the evolution of the spatial galaxy number density and stellar mass density distributions are a key to understand how stellar mass accretes onto massive haloes. We perform these measurements in GCLASS by dividing the sample in radial bins. We do this by stacking the cluster ensemble at the location of the BCGs, and scaling the clusters by their respective $R_{200}$. We measure the area in each bin by masking the locations on the images that are contaminated by bright stars. Also, since we do not take the stellar mass of the BCGs into account in this study, we mask the location of the central galaxies since this location does not allow for the detection of typical cluster members.

\begin{figure}
\resizebox{\hsize}{!}{\includegraphics{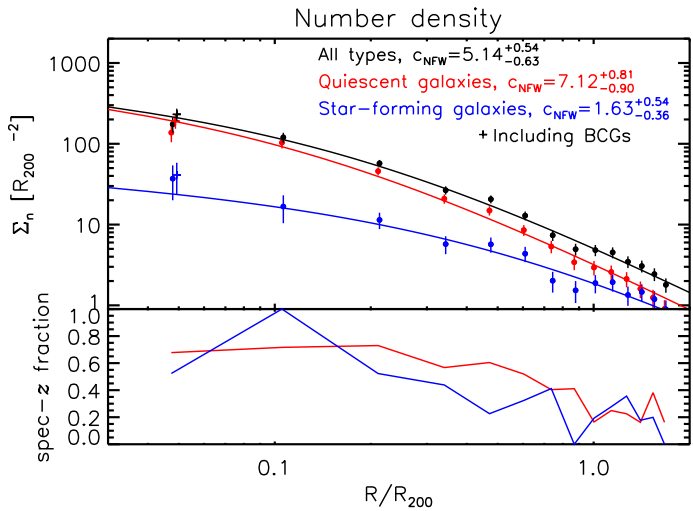}}
\caption{The number density of galaxies with stellar masses $>10^{10.2}\,\rm{M_{\odot}}$ in the 10 GCLASS clusters as a function of radial distance. The total galaxy population (black) is separated between star-forming (blue) and quiescent (red) galaxies. Thick points show the membership-corrected number density, where the error bars represent the uncertainties that arise from membership correction. The points are fitted by projected NFW functions (lines), with different concentration parameters. The lower panel shows the fraction of galaxies in each bin with a spectroscopic redshift.}
\label{fig:radnumberdens}
\end{figure}

The number density distribution is shown in Fig.~\ref{fig:radnumberdens}, where in each radial bin the number of spec-$z$ identified cluster members and the membership-corrected photo-$z$ members with stellar masses exceeding $10^{10.2}\,M_{\odot}$ are combined. 
Errors on each point are a combination of Poisson sampling errors, and errors propagated from the membership correction which we estimated from a series of Monte-Carlo simulations. We used the area-weighted position to plot the data points in the horizontal direction. The $^\backprime+^\prime$-signs show the innermost point including the BCGs. The bottom panel of Fig.~\ref{fig:radnumberdens} shows the spectroscopic targeting completeness as a function of radial distance, which shows that -as designed- the completeness is higher for objects near the projected cluster centres. Further away from the cluster centre, the errors that arise from membership estimates are dominant.

\begin{figure}
\resizebox{\hsize}{!}{\includegraphics{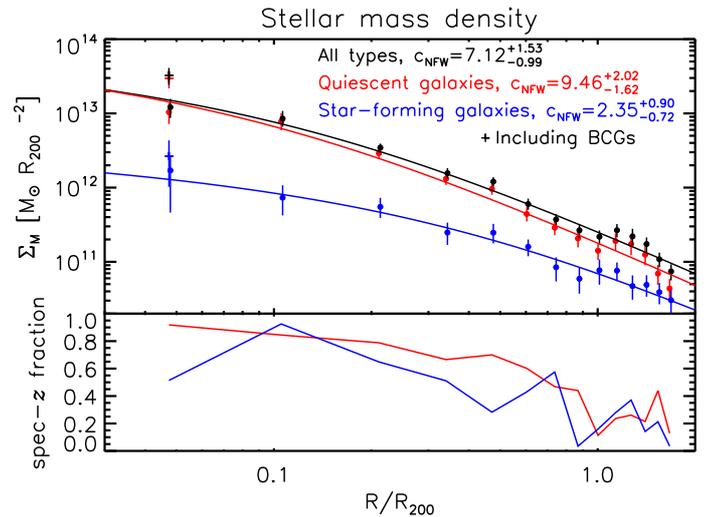}}
\caption{The stellar mass density distribution in galaxies with stellar masses $>10^{10.2}\,\rm{M_{\odot}}$ of the composite cluster as a function of radial distance. Comparing these distributions to those shown in Fig.~\ref{fig:radnumberdens}, we find that the stellar mass distributions are peaked more strongly than the number density distribution. That is an indication for mass segregation of quiescent galaxies in these systems. The lower panel shows the fraction of stellar mass in galaxies with a spectroscopic redshift.}
\label{fig:stelmassdens}
\end{figure}

The radial distribution of stellar mass in the ensemble cluster is shown in Fig.~\ref{fig:stelmassdens}. Besides Poisson counting errors and errors that arise from cluster membership corrections, the error bars include stellar mass measurement errors on individual galaxies. Compared to the number density profile, the spectroscopic targeting completeness is higher due to the selection of spectroscopic targets by their 3.6$\mu$m flux.

\begin{table}[h]
\caption{Best-fitting NFW parameters to the radial density distributions. Reduced $\chi^2$ values are given (14 degrees of freedom).}
\large
\label{tab:nfwpars}
\begin{center}
\begin{tabular}{l c c}
\hline
\hline
&$c_{\rm{NFW}}$&$\chi^2/d.o.f.$\\
\hline
$\Sigma_{\rm{M,all}}$&$  7.12^{+  1.53}_{-  0.99}$&  0.94\\
$\Sigma_{\rm{n,all}}$&$  5.14^{+  0.54}_{-  0.63}$&  0.84\\
$\Sigma_{\rm{M,quiescent}}$&$  9.46^{+  2.02}_{-  1.62}$&  1.07\\
$\Sigma_{\rm{n,quiescent}}$&$  7.12^{+  0.81}_{-  0.90}$&  0.92\\
$\Sigma_{\rm{M,star-forming}}$&$  2.35^{+  0.90}_{-  0.72}$&  0.36\\
$\Sigma_{\rm{n,star-forming}}$&$  1.63^{+  0.54}_{-  0.36}$&  0.73\\
\hline
\end{tabular}
\end{center}
\end{table}

We fit projected NFW \citep{NFW} profiles to the data points, excluding the BCGs, to be able to interpret the results in the context of the NFW concentration parameter. Using $\chi ^2$ minimization, taking account of the 2D annulus-shaped bins, we find the best fitting functions, which give good representations of the data (see the reduced $\chi^2$ values in Table~\ref{tab:nfwpars}). We give the best-fitting concentration parameters and their marginalized errors in the table for both the number density and the stellar mass density profiles. The best-fitting profiles are shown in the corresponding figures. 

From both the number density and the stellar mass density profiles we find that the quiescent galaxy population is concentrated more strongly than the star-forming population, which is consistent with the view that the star-forming population is accreted more recently by the cluster \citep[for a measurement at low-$z$, cf.][]{biviano02}. 

We also find that the stellar-mass distribution of quiescent galaxies is concentrated more strongly than their number density profile, which is an indication that more massive galaxies are situated closer towards the cluster centres than lower mass galaxies. This is likely caused by dynamical friction of the cluster members, which is more efficient for massive galaxies. Note that this effect is observed without taking account of the BCGs.

\subsection{Discussion}\label{sec:discussion}
We measured the galaxy concentration parameters in the ensemble GCLASS cluster, and it may be that a subset of these systems is driving the concentration to this relatively high value. To investigate this we perform different stacks using subsets of the GCLASS sample. We separate the sample in 3 bins, and to make sure the statistics in each bin are sufficiently high, we rank order the clusters by total stellar mass and fill the bins by 6, 3, and 1 cluster(s), respectively. We find that the best-fitting stellar mass concentrations for these 3 ensembles are in the range $6.0<c<9.0$, and agree to within $2\sigma$ of their measurement errors. This suggests that the stellar mass in each of the GCLASS clusters is likely to be distributed with a concentration parameter around $c\sim 7$.

This high concentration parameter for the stellar matter suggest that the stellar mass is concentrated more strongly than the dark matter is expected to be. For the GCLASS haloes \citet{duffy08} estimates a concentration parameter around $c=2.7$ from simulations that only contain dark matter. Although this value is the median value for massive haloes at $z=1$, the distribution of concentrations is found to be distributed by a log-normal distribution with a scatter $\sigma(\log(c))=0.15$. It is possible that the red-sequence selection method is biased towards systems with highly concentrated red-sequence galaxies. However, given the large difference in concentration between the stellar mass and dark matter, and the relatively small scatter in the distributions, it is unlikely that this difference is merely an effect of the selection method. Note that it is possible that the inclusion of baryonic physics in simulations will alter the dark matter distribution, as recent studies have suggested \citep[e.g.][]{vandaalen11}. This might bring the dark matter and stellar mass concentrations better in agreement. We checked that the results shown in Fig.~\ref{fig:stellmassvshalomass} are only marginally affected if we change the concentration to $c=7$.

The composite cluster sample is obtained after stacking the individual clusters on the locations of their BCGs. In some cases the identification of the BCG is ambiguous. For SpARCS-1051 and SpARCS-1634 the identified BCGs are separated by $\sim 250$kpc from the approximate projected cluster centres. We test what the effect of possible mis-centring is on the concentration of the measured radial density profiles. We find that, if the intrinsic cluster profiles are described by a $c=10$ NFW profile, and 10 clusters are stacked with a mis-centring sampled from a Gaussian distribution with $\sigma=0.1 r_{200}$, the measured concentration would be $c=7$. Any misalignment with the "true" cluster centre would result in a concentration that is biased low. Given these tests, it is likely that the stellar mass is concentrated even more strongly than indicated by the NFW fits to the cluster ensemble.

\subsection{Evolution towards lower redshift}
From numerical simulations \citep{wechsler02} we know that massive haloes are likely to grow by a factor of $\sim2.5$ between $z=1.0$ and $z=0.3$. This suggests that the GCLASS cluster sample, with typical halo masses of $\rm{M_{200}} \simeq 2 \times 10^{14}$, is the likely progenitor population of the clusters observed in the CNOC survey \citep{yee96,carlberg96}, which have typical halo masses of $\rm{M_{200}} \simeq 7 \times 10^{14}$. The concentration of the underlying dark matter distribution is expected to increase by $\sim10\%$ in this redshift interval \citep{duffy08}. \citet{muzzin07} measured the K-band luminosity and number density profiles for 15 of the CNOC1 clusters, and showed that the K-band luminosity distribution is well described by a projected NFW profile with concentration parameter $c=4.28\pm0.57$. Although the luminosity in the K-band is a good proxy for the stellar mass, the mass-to-light ratio in this filter depends on galaxy type. Since we find a different distribution of stellar mass in quiescent and star-forming galaxies (Fig.~\ref{fig:stelmassdens}), this suggests that the K-band luminosity profile differs from the stellar mass density profile. Indeed, if we scale the star-forming galaxies in GCLASS by a factor of 2 to account for the rough difference in $M_{\star}/L$, we measure a luminosity profile with a concentration $c<6$. Although the difference between GCLASS and CNOC1 is hence not as extreme, these results suggest that the dark matter and stellar mass density distributions evolve in distinct ways. This is also suggested by \citet{budzynski12}, who based their study on a sample of groups and clusters in the redshift range $0.15<z<0.4$ from the SDSS. For this sample \citet{budzynski12} found that the concentration of the number density profile is lower than the dark matter prediction. There are several caveats, and possible explanations for the observed evolution of the stellar mass distribution.

First, since we do not take account of the stellar mass present in the central galaxies when fitting NFW profiles, accretion of galaxies onto the central galaxy might change the distribution of stellar mass in satellites, and therefore the concentration parameter, over time. Mergers play a dominant role in the build-up of stellar mass in BCGs \citep{lidman13,burke13}. Massive galaxies that are close to the centre are expected to merge with the BCG on a relatively short timescale \citep{bildfell12,lidman13}, thereby rendering the BCG an increasingly statistically different population compared to cluster satellite galaxies. An indicator for this process is an increase of the luminosity gap between the BCG and the second brightest cluster galaxy \citep[e.g.][]{smith10}. However, given the shallow slope of the central stellar-halo mass relation (Sect.~\ref{sec:bcgmasshalomass}), BCGs are expected to grow only by a factor of 1.5 in the redshift range $1.0>z>0.3$ (see also \citet{lidman12}). If the supply of this stellar mass growth is obtained from galaxies near the centre, the concentration parameter of the satellite galaxy population would go down. 

We perform a simple simulation in which we reduce the stellar mass in satellite galaxies within $0.5\cdot R_{200}$ in accordance with a BCG growth of a factor of 1.5, and this shows that this is not sufficient to explain the dramatic decrease in the concentration parameter ($c$ decreases from 7.0 to 6.0). Nevertheless, it is possible that the build-up of the ICL component towards lower redshift plays a role in lowering the concentration parameter of stellar mass in satellites. 

Second, as clusters get larger, the dynamical friction timescale of a galaxy with a given mass increases, so that it takes longer for galaxies to sink to the centre of the potential well. This is also hinted at when we compare the relation between central stellar and halo mass (Sect.~\ref{sec:bcgmasshalomass}), and between total stellar mass and halo mass (Sect.~\ref{sec:results1}). Given that the latter slope is steeper, the fraction of stellar mass in satellite galaxies is higher in more massive haloes. It is possible that galaxies that are accreted onto the cluster at a later time are situated closer to the outskirts of the clusters due to the same process, and thus are less concentrated than the population that was accreted earlier.

We perform a simple test in which we increase the mass of the ensemble cluster by a factor 2.5 by adding stellar mass that is distributed following an NFW distribution with a given concentration. If we vary the concentration parameter of the population that we add, we find that, in order to end up with a concentration of $c=4.0$ by $z=0.3$ (i.e., similar to the concentration measured in CNOC), we have to add satellites with a concentration parameter of $c=2.8$ to the stellar mass density distribution observed in GCLASS. This scenario could potentially explain the difference with the results from \citet{budzynski12}, who find that at low-$z$ the stellar mass is concentrated more strongly than the dark matter, and suggests that the stellar mass content mostly grows by accreting stellar mass onto the cluster outskirts. 

\section{Summary and Conclusions}\label{sec:conclusions}
In this paper we provide three key measurements concerning the stellar content in 10 clusters at $z\sim1$ from the GCLASS survey. GCLASS benefits from 11 band photometric coverage and deep spectroscopic coverage to provide a full census of stellar mass in cluster members down to about $M_{\star}=10^{10.2}\,\rm{M_{\odot}}$. Combining these observations with measurements at lower redshifts we hope to provide constraints on the way baryons cool and form stars in galaxies in high density environments.

In Sect.~\ref{sec:bcgmasshalomass} we presented a comparison of the central stellar mass with total halo mass, and found a correlation that suggests that the fraction of mass in the central galaxy is a decreasing function of halo mass, and about 0.001 for the mass range probed by GCLASS. We confirmed the trend predicted using abundance matching techniques, both in a qualitative as quantitative sense. 

Sect.~\ref{sec:results1} showed a comparison of the total stellar masses (including satellites) with the dynamical halo masses, both within $R_{200}$ and $R_{500}$. We found that the total stellar mass increases with halo mass, and that the fraction is around 0.01 for our sample and appears to decrease towards higher halo masses. A comparison of this relation with samples at other redshifts can yield insights on the way these systems accrete their stellar mass, but is difficult due to inhomogeneous sample selections and analyses. Especially inaccurate estimates on the stellar mass-to-light ratio are a source of confusion. After correcting the reference studies for differences in their analyses, we found no significant evolution with redshift in the stellar mass fraction at fixed halo mass.

In Sect.~\ref{sec:results2} we studied the radial number density and stellar mass density profiles of galaxies in the sample, and found that these are represented by projected NFW profiles. The stellar mass density distribution is concentrated more strongly than the galaxy number density distribution, which shows that more massive galaxies are situated closer to the cluster cores (i.e. mass segregation). The stellar mass density profile has an NFW concentration parameter ($c=7$) that is significantly higher than the dark matter distribution is expected to be ($c=2.7$) from numerical simulations. Comparison of the concentration parameter with the CNOC1 survey at $z=0.3$ suggests that the stellar mass concentration should decrease towards lower redshift. A simple simulation showed that stellar mass growth of the BCG alone is not enough to explain the evolution between GCLASS and CNOC1, and that the clusters are likely to accrete more stellar mass on the cluster outskirts as they grow by a factor of 2.5 in total mass from $z=1$ to $z=0.3$. Also the build-up of the ICL can play a role in the observed evolution.

We note that comparisons of our results with other studies are complicated due to inhomogeneous samples and different analyses. In order to draw firm conclusions regarding the evolution of the baryonic content, and in particular stellar mass, observational data need to be homogenized. We have also seen that the assumption of a single stellar mass-to-light ratio is inadequate to measure the total stellar mass content of galaxy clusters. Rather, one should fit the full Spectral Energy Distributions (SEDs) to estimate the stellar masses for individual galaxies. Moreover, since the galaxies with a high mass-to-light ratio are generally more concentrated in the cluster cores, measurements of the K-band luminosity profile and stellar mass density profile should not be taken as equivalent measurements. Thanks to the advance of large optical and near-infrared imaging facilities over the past decades, these multi-wavelength data are relatively easy to obtain, so that we will soon expect to be able to compare consistent stellar mass measurements with full SED fitting over a large redshift baseline.

\begin{acknowledgements}
We thank Marcello Cacciato, Dennis Just, Rob Crain, Anthony Gonzalez, Michael Balogh and Ivo Labb\'e for valuable discussions that improved the quality of this work.

R.F.J. van der Burg and H. Hoekstra acknowledge support from the Netherlands Organisation for Scientic Research grant number 639.042.814. C. Lidman is the recipient of an Australian Research Council Future Fellowship (program number FT0992259).

Based on data products from observations made with ESO Telescopes at the La Silla Paranal Observatories
under ESO programme ID 179.A-2005 and on data products produced by TERAPIX and the Cambridge Astronomy
Survey Unit on behalf of the UltraVISTA consortium.
\end{acknowledgements}

\bibliographystyle{aa} 
\bibliography{MasterRefs} 

\end{document}